\documentclass[aps,amsmath,amssymb, prc,twocolumn,superscriptaddress]{revtex4-1}

\usepackage{graphicx}
\usepackage{dcolumn}
\usepackage{bm}
\usepackage[utf8]{inputenc}
\usepackage[ngerman,english]{babel}
\usepackage{float}
\usepackage{multirow}
\usepackage{longtable}
\usepackage{dcolumn}
\newcolumntype{d}{D{.}{.}{-1}}
\usepackage{setspace}
\usepackage{threeparttable}
\usepackage{tabularx}
\usepackage[symbol*]{footmisc}
\DefineFNsymbolsTM{myfnsymbols}{
  \textasteriskcentered *
  \textdagger    \dagger
  \textdaggerdbl \ddagger
  \textsection   \mathsection
  \textbardbl    \|%
  \textparagraph \mathparagraph
}%
\setfnsymbol{myfnsymbols}

\usepackage{siunitx}
\newcommand{\nuc}[2]{\hbox{$^{#1}$#2}}

\begin{document}


\title{One-proton and one-neutron knockout reactions from $N=Z=28$ $^{56}$Ni to
  the $A=55$ mirror pair $^{55}$Co and $^{55}$Ni}


\author{M. Spieker}
\email[]{spieker@nscl.msu.edu}
\affiliation{National Superconducting Cyclotron Laboratory, Michigan State University, East Lansing, Michigan 48824, USA}

\author{A. Gade}
\affiliation{National Superconducting Cyclotron Laboratory, Michigan State University, East Lansing, Michigan 48824, USA}
\affiliation{Department of Physics and Astronomy, Michigan State University, East Lansing, Michigan 48824, USA}

\author{D. Weisshaar}
\affiliation{National Superconducting Cyclotron Laboratory, Michigan State University, East Lansing, Michigan 48824, USA}

\author{B.\ A. Brown}
\affiliation{National Superconducting Cyclotron Laboratory, Michigan State University, East Lansing, Michigan 48824, USA}
\affiliation{Department of Physics and Astronomy, Michigan State University, East Lansing, Michigan 48824, USA}

\author{J.\ A. Tostevin}
\affiliation{Department of Physics, Faculty of Engineering and Physical Sciences, University of Surrey, Guildford, Surrey GU2 7XH, United Kingdom}

\author{B. Longfellow}
\affiliation{National Superconducting Cyclotron Laboratory, Michigan State University, East Lansing, Michigan 48824, USA}
\affiliation{Department of Physics and Astronomy, Michigan State University, East Lansing, Michigan 48824, USA}

\author{P. Adrich}
\altaffiliation{Present address: National Centre for Nuclear Research, Otwock, Poland}
\affiliation{National Superconducting Cyclotron Laboratory, Michigan State University, East Lansing, Michigan 48824, USA}

\author{D. Bazin}
\affiliation{National Superconducting Cyclotron Laboratory, Michigan State University, East Lansing, Michigan 48824, USA}
\affiliation{Department of Physics and Astronomy, Michigan State University, East Lansing, Michigan 48824, USA}

\author{M.\ A. Bentley}
\affiliation{Department of Physics, University of York, Heslington, York YO10 5DD, United Kingdom}

\author{J.\ R. Brown}
\affiliation{Department of Physics, University of York, Heslington, York YO10 5DD, United Kingdom}

\author{C.\ M. Campbell}
\altaffiliation{Present address: Nuclear Science Division, Lawrence Berkeley National Laboratory, Berkeley, California 94720, USA}
\affiliation{National Superconducting Cyclotron Laboratory, Michigan State University, East Lansing, Michigan 48824, USA}
\affiliation{Department of Physics and Astronomy, Michigan State University, East Lansing, Michigan 48824, USA}

\author{C.\ Aa. Diget}
\affiliation{Department of Physics, University of York, Heslington, York YO10 5DD, United Kingdom}

\author{B. Elman}
\affiliation{National Superconducting Cyclotron Laboratory, Michigan State University, East Lansing, Michigan 48824, USA}
\affiliation{Department of Physics and Astronomy, Michigan State University, East Lansing, Michigan 48824, USA}

\author{T. Glasmacher}
\affiliation{National Superconducting Cyclotron Laboratory, Michigan State University, East Lansing, Michigan 48824, USA}
\affiliation{Department of Physics and Astronomy, Michigan State University, East Lansing, Michigan 48824, USA}

\author{M. Hill}
\affiliation{National Superconducting Cyclotron Laboratory, Michigan State University, East Lansing, Michigan 48824, USA}
\affiliation{Department of Physics and Astronomy, Michigan State University, East Lansing, Michigan 48824, USA}

\author{B. Pritychenko}
\affiliation{National Nuclear Data Center, Brookhaven National Laboratory, Upton, New York 11973, USA}

\author{A. Ratkiewicz}
\altaffiliation{Present address: Lawrence Livermore National Laboratory, Livermore, California, 94550, USA}
\affiliation{National Superconducting Cyclotron Laboratory, Michigan State University, East Lansing, Michigan 48824, USA}
\affiliation{Department of Physics and Astronomy, Michigan State University, East Lansing, Michigan 48824, USA}

\author{D. Rhodes}
\affiliation{National Superconducting Cyclotron Laboratory, Michigan State University, East Lansing, Michigan 48824, USA}
\affiliation{Department of Physics and Astronomy, Michigan State University, East Lansing, Michigan 48824, USA}


\date{\today}

\begin{abstract}

We present a high-resolution in-beam $\gamma$-ray spectroscopy study of excited
states in the mirror nuclei $^{55}$Co and $^{55}$Ni following one-nucleon
knockout from a projectile beam of $^{56}$Ni. The newly determined partial cross
sections and the $\gamma$-decay properties of excited states provide a test of
state-of-the-art nuclear structure models and probe mirror symmetry in 
unique ways. The new experimental data are compared to large-scale
shell-model calculations in the full $pf$ space which include charge-dependent
contributions. A mirror asymmetry for the partial cross sections leading to the
two lowest $3/2^-$ states in the $A=55$ mirror pair was identified as well as a
significant difference in the $E1$ decays from the $1/2^+_1$ state to the same
two $3/2^-$ states. The mirror asymmetry
in the partial cross sections cannot be reconciled with the present shell-model picture or small mixing introduced in a two-state model. The observed mirror asymmetry in the $E1$ decay pattern, however, points at stronger mixing between the two lowest $3/2^-$ states in $^{55}$Co than in its mirror $^{55}$Ni.
\end{abstract}


\maketitle


The concept of isospin symmetry in atomic nuclei is rooted in the fundamental assumption of charge symmetry and charge independence of the attractive nucleon-nucleon interaction, see the review article\,\cite{War06}. In the absence of isospin-breaking effects, such as the Coulomb force, an exact degeneracy of isobaric analog states (IAS) with isospin quantum number $T$ in nuclei of same mass but with interchanged neutron and proton numbers (mirror pairs) would be expected. Thus, observed differences of IAS properties in mirror nuclei can elucidate the presence and nature of isospin-breaking contributions to the nuclear many-body problem. Excitation-energy shifts between mirror pairs, so-called mirror energy differences (MED), were systematically studied to identify such contributions in the $pf$ shell\,\cite{Ben07a, Ben15}, i.e. for nuclei between the doubly-magic $N=Z$ nuclei $^{40}$Ca and $^{56}$Ni. Unexpected asymmetries in the $E1$ decay pattern of low-lying excited states were also observed between mirror pairs\,\cite{Ekm04, Jen05, Ben06, Orl09}. Their origin has been traced back to isospin-symmetry violation though the exact underlying mechanism is still discussed\,\cite{Pat08, Biz12}. 

We report on a study that uses mirrored one-neutron and one-proton knockout
reactions from \nuc{56}{Ni} to the mirror nuclei
\nuc{55}{Ni} and \nuc{55}{Co}, respectively. Similar types of mirrored reactions
have been employed before to extract MED in more distant mirror
pairs such as (\nuc{52}{Ni},\nuc{52}{Cr})\,\cite{Dav13}, (\nuc{53}{Ni},\nuc{53}{Mn})\,\cite{Mil16}
and (\nuc{70}{Se},\nuc{70}{Kr})\,\cite{Wim18a}, however, starting from projectiles
that are mirrors themselves rather than from a self-conjugate
nucleus. Brown {\it et al.}\,\cite{Bro09} used the $\gamma$-ray spectra of
\nuc{53}{Ni} and \nuc{53}{Mn} from the three-neutron and three-proton removal on
\nuc{56}{Ni} projectiles to match analog states but such reactions cannot be
described within a direct reaction formalism.   

The doubly-magic nucleus \nuc{56}{Ni} and the $T = 1/2$ ($T_z = \pm 1/2$) (\nuc{55}{Co},\nuc{55}{Ni}) mirror pair are of particular interest as they are coming within reach of ab-initio-type calculations ~\cite{Hag16,Her17} that compute nuclei based on
forces from chiral effective field theory. \nuc{56}{Ni} has also been a target
for early large-scale configuration-interaction shell-model calculations in the
full $pf$ model space\,\cite{Hor06}, pioneering coupled-cluster
calculations\,\cite{Hor07,Gou08}, and self-consistent Green's function
theory\,\cite{Bar09a, Bar09b}. Although nominally doubly-magic, \nuc{56}{Ni} behaves as a soft core
in shell-model calculations performed in the full $pf$ model space. Using the
effective isospin-conserving GXPF1A interaction, the closed-shell $(1f_{7/2})^{16}$
configuration comprises only about 68\,$\%$ of the ground-state
wavefunction\,\cite{Hon04}. These calculations successfully account for the
observed quadrupole collectivity~\cite{Yur04b} and the ground-state magnetic moments of the
odd-$A$ neighbors with one nucleon added or
removed\,\cite{Cal73, Oht96, Ber09, Coc09}. 

In terms of single-particle properties, a number of
experiments\,\cite{ENSDF, Sho72, For77, Erl77, Zho96, Yur06, Jia09, Lee09,
  Sch13, San14} and theoretical studies\,\cite{Tra96, Hon04, Hor06, Hor07,
  Bar09a, Bar09b, Lit11} have been performed to identify the fragments of the
single-particle levels relative to the $N=Z=28$ core and have also suggested the existence of $(2^+_1({}^{56,58}\mathrm{Ni}) \otimes 1f_{7/2}^{-1})$ core-coupled excitations with $J^{\pi} = 3/2^-, ..., 11/2^-$ in
the vicinity of $^{56}$Ni\,\cite{Bur71, Bro87, Yur04}. Only recently, an
inverse-kinematics one-neutron transfer experiment
$^{1}\mathrm{H}({}^{56}\mathrm{Ni},d){}^{55}\mathrm{Ni}$ populated for the first
time single-hole-like states directly from the $^{56}$Ni ground
state\,\cite{San14}, however, without detecting subsequent $\gamma$-ray emission. An excited $3/2^-$ and $1/2^+$ state were observed. 

Relevant for this work, strong isospin mixing between the $T = 3/2$, $J^{\pi} = 3/2^-$ IAS of $^{55}$Cu and a very close-lying $T =
1/2, J^{\pi} = 3/2^-$ state was observed in a $\beta$-decay
experiment leading to $^{55}$Ni\,\cite{Tri13}. Very similar observations had been
made in the 1970s for the IAS of $^{55}$Fe in the mirror nucleus
$^{55}$Co\,\cite{Sho72, Mar76, For77}. The recent $\beta$-decay data on $^{55}$Ni hint at
slightly stronger isospin mixing in $^{55}$Co\,\cite{Tri13} as compared to the
earlier work mentioned above. The degree of isospin mixing between $T=0$
and $T=1$ components in the ground state of $^{56}$Ni has been controversially
discussed. Some evidence comes from the detection of $\beta$-delayed protons
after the $\beta^+$ decay of $^{57}$Zn ($T=3/2$), where both the $0^+_1$ and
$2^+_1$ state in $^{56}$Ni ($T=0$) were strongly populated\,\cite{Jok02}. 

 Here, we investigate the single-particle structure of self-conjugate
 \nuc{56}{Ni} and the mirrors \nuc{55}{Ni} (\nuc{55}{Co}) using the $\gamma$-ray-tagged mirrored one-neutron (one-proton) knockout reactions. Mirror asymmetries in partial cross sections and $\gamma$-decay patterns will be discussed.  

\begin{figure}[t]
\centering
\includegraphics[width=1\linewidth]{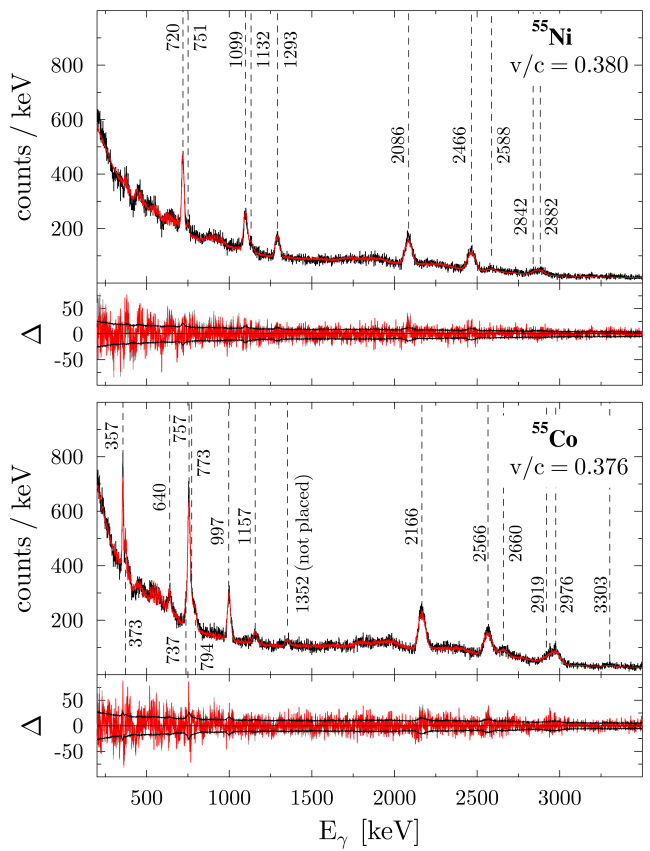}
\caption{\label{fig:spectrum}{(color online) In-beam $\gamma$-ray singles
    spectra for $^{55}$Ni (top) and $^{55}$Co (bottom) in black compared to
    $\gamma$-ray spectra obtained from a GEANT4 simulation (red). Observed
    transitions are marked with dashed vertical lines and their corresponding
    transition energies. Also shown are the fit residuals $\Delta$ (red) in
    combination with the $1\sigma$ confidence level (black) in the lower
    panels. The background structures between $400-800$\,keV, seen on top of the smooth background, are caused by $\gamma$ rays emitted from stopped components and taken into account in the simulation.}} 
\end{figure}

The experiment was performed at the Coupled Cyclotron Facility of the National
Superconducting Cyclotron Laboratory (NSCL) at Michigan State
University\,\cite{NSCL}. The secondary beam of $^{56}$Ni was selected in flight
with the A1900 fragment separator\,\cite{Mor03} using a 300 mg/cm$^2$ Al
degrader after production from a 160\,MeV/u $^{58}$Ni primary beam in projectile
fragmentation on a thick 610 mg/cm$^2$ $^{9}$Be target. The $^{56}$Ni secondary
beam was unambigiously distinguished from 
the $^{55}$Co (27\,$\%$) and $^{54}$Fe\,(1\,$\%$) contaminants via the
time-of-flight difference measured between two plastic scintillators located at
the exit of the A1900 and the object position of the S800 analysis beam
line. A 188 mg/cm$^2$ $^{9}$Be reaction target was surrounded by
the SeGA array consisting of 16 32-fold segmented High-Purity Germanium
detectors\,\cite{Mue01}. The detectors were arranged in two rings with central
angles of 37$^{\circ}$ (7 detectors) and 90$^{\circ}$ (9 detectors) relative to
the beam axis. The segmentation of the detectors enables an event-by-event
Doppler reconstruction of the $\gamma$ rays emitted by the projectile-like
reaction residues in flight ($v/c \approx 0.4$). The angle of the $\gamma$-ray
emission needed for this reconstruction is determined from the segment position
that registered the highest energy deposition. All projectile-like reaction
residues entering the S800 focal plane were identified event-by-event from their
energy loss and time of flight\,\cite{Baz03}. The Doppler-corrected in-beam
$\gamma$-ray singles spectra in coincidence with event-by-event identified
knockout residues are shown in Fig.\,\ref{fig:spectrum}. Only a change in
magnetic rigidity of the S800 spectrograph was required to switch from one knockout setting to the
other.

Inclusive cross sections of $38.0 \pm 0.2 \, (\mathrm{stat.}) \pm 3.0 \,
(\mathrm{sys.})$\,mb for the one-neutron knockout from $^{56}$Ni to all bound states
of $^{55}$Ni and of $126 \pm 2 \, (\mathrm{stat.}) \pm 17 \,
(\mathrm{sys.})$\,mb for the one-proton knockout to all bound final states of
$^{55}$Co were determined. In both cases, the inclusive cross
section was deduced from the yield of the detected knockout residues relative to
the number of incoming $^{56}$Ni projectiles and the number density of
the $^{9}$Be reaction target. Statistical and systematic uncertainties are
quoted separately. The latter 
include the stability of the secondary beam composition, the choice of software
gates, and corrections for acceptance losses in the tails of the residue
parallel momentum distributions due to the blocking of the unreacted beam in the focal plane. The parallel 
momentum distribution of the knockout residues was reconstructed on an
event-by-event basis using the two position-sensitive cathode readout drift
counters of the S800 focal-plane detection system\,\cite{Baz03} in
conjunction with trajectory reconstruction through the spectrograph.

\begin{figure}[t]
\centering
\includegraphics[width=1\linewidth]{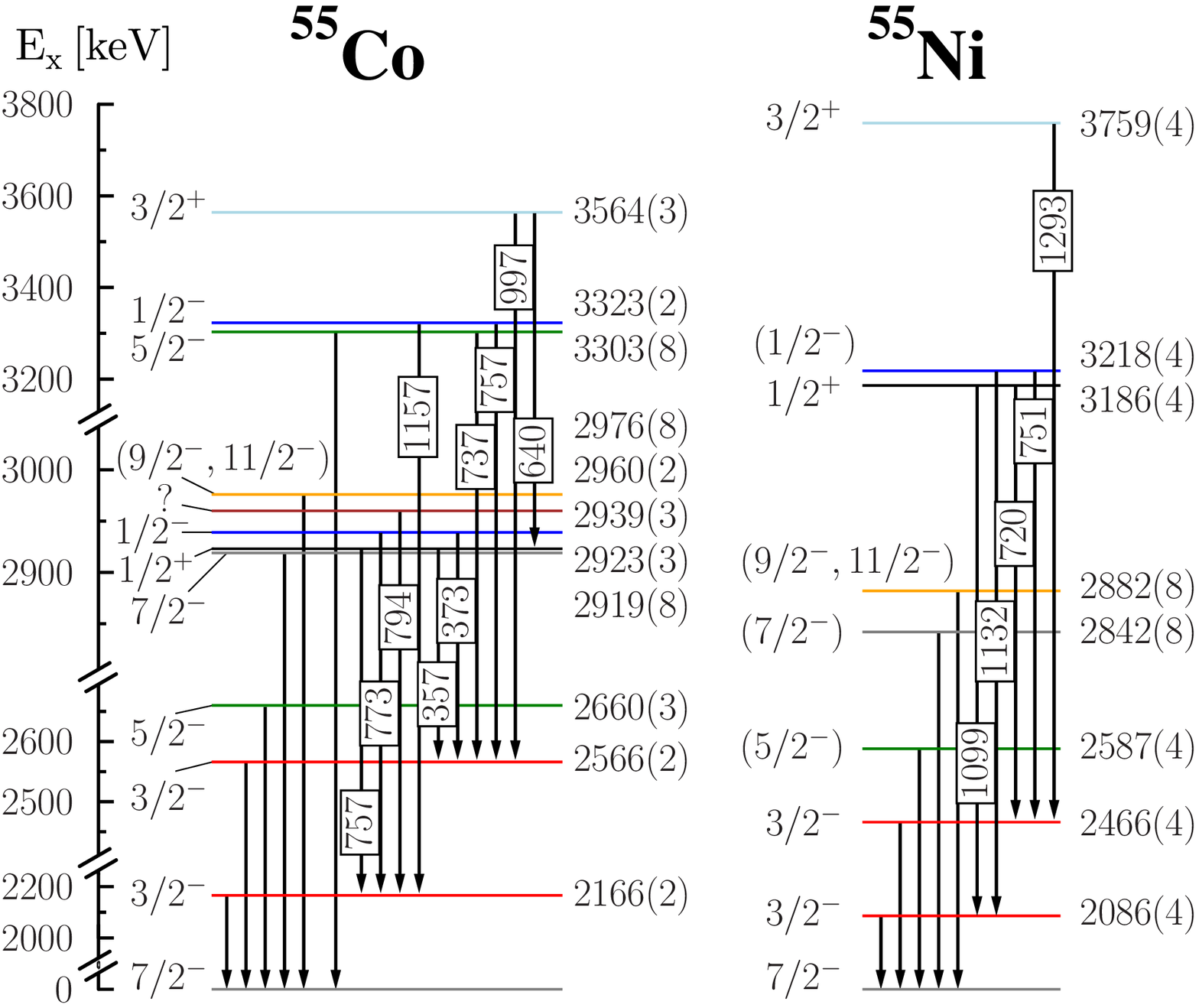}
\includegraphics[width=1\linewidth]{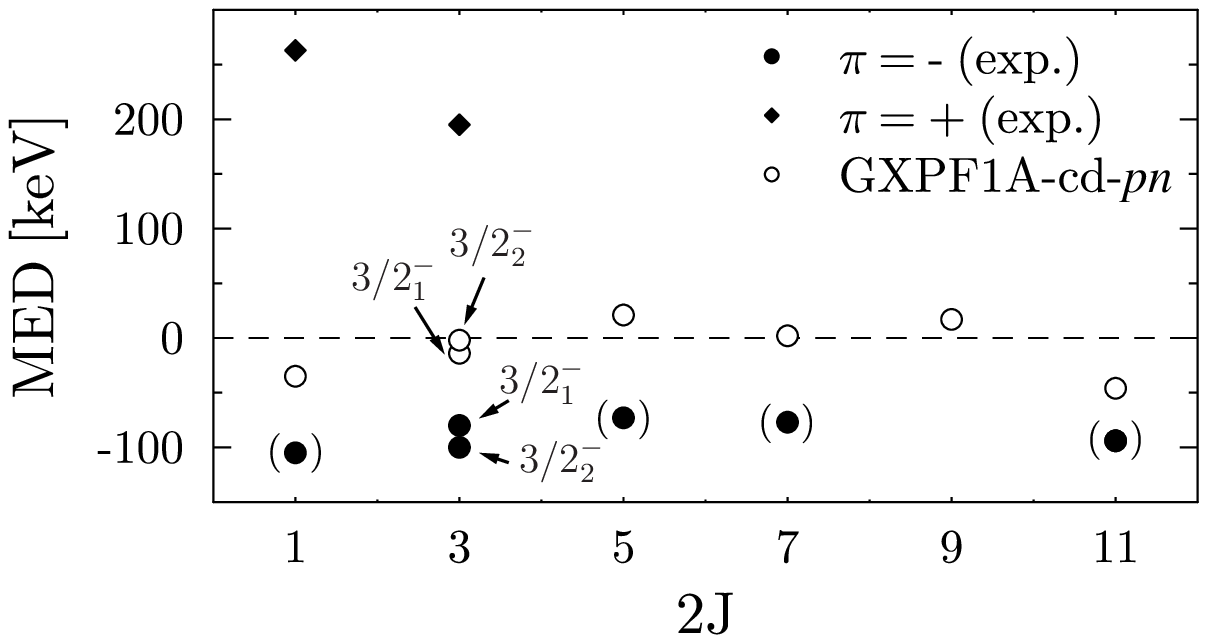}
\caption{\label{fig:level}{(color online) Level scheme observed for $^{55}$Co
    and $^{55}$Ni. All transitions visible in Fig.\,\ref{fig:spectrum} with the exception of the 1352\,keV ($^{55}$Co) transition are
    placed. The color coding is the same as in
    Fig.\,\ref{fig:partials}. The bottom panel shows the calculated MED for the two $3/2^-$, $5/2^-_1$, $1/2^-_2$, $7/2^-_2$, 
    $(9/2^-,11/2^-)$ (solid circles) as well as the $1/2^+$ and $3/2^+$ states (solid diamonds) in comparison to the shell-model results (open circles, $pf$ states only). Even though the transition intensities seen in Fig.\,\ref{fig:spectrum} are
    comparable, the MED for the excited $J^{\pi} = 5/2^-_1, 1/2^-_2, 7/2^-_2$ and
    $(9/2^-,11/2^-)$ states are only tentatively assigned and, thus, shown in parentheses. Despite the $3/2^+$ state discussed in the text, $J^{\pi}$ assignments were adopted from Refs.\,\cite{ENSDF, Tri13, San14}.}} 
\end{figure}

To calculate the $\gamma$-ray yields needed to determine the partial cross
sections to individual final states, GEANT4 simulations were performed
with the UCSeGA simulation package\,\cite{ucsega}. The results of those
simulations, assuming a smooth double-exponential background, are shown in
Fig.\,\ref{fig:spectrum} together with the measured $\gamma$-ray spectra. Possible sources of the in-beam background were discussed in, {\it e.g.},\,\cite{Str14, Pod03, Wei08}. Using $\gamma\gamma$ coincidences, feeders were
identified and the placement of previously known $\gamma$-ray transitions in the level scheme\,\cite{ENSDF, Tri13}
confirmed. The level schemes are displayed in Fig.\,\ref{fig:level}. 

\begin{figure}[t]
\centering
\includegraphics[width=1\linewidth]{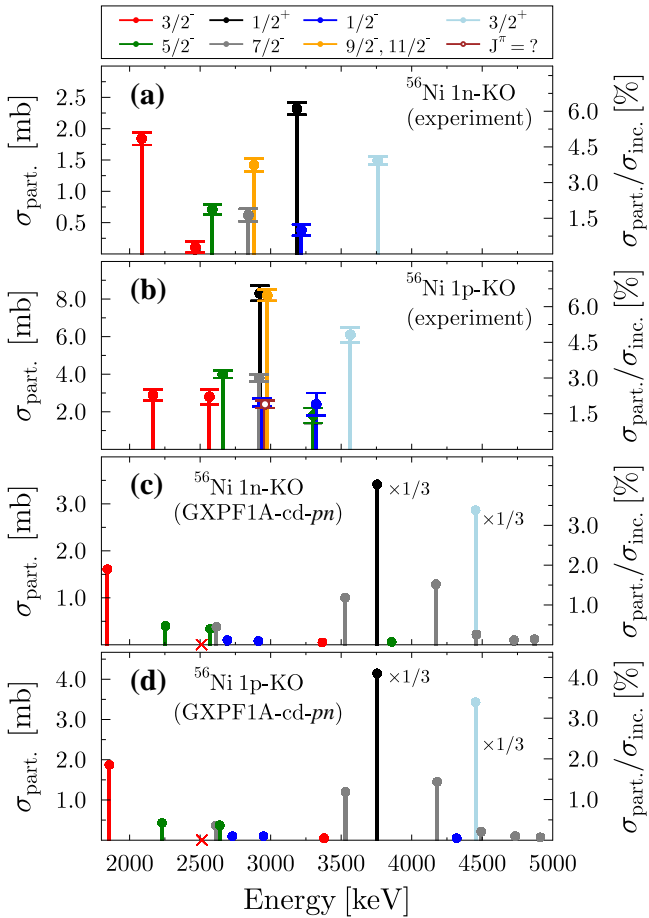}
\caption{\label{fig:partials}{(color online) Partial cross sections
    $\sigma_{\mathrm{part.}}$ determined for {\bf (a)} $^{55}$Ni and {\bf (b)}
    $^{55}$Co in comparison to {\bf (c), (d)} the theoretical cross
    sections. Only states predicted with $\sigma_{part.} \geq 0.05$\,mb are
    presented. The spectroscopic factors and excitation energies predicted for
    the $1/2^+$ and $3/2^+$ state have been been taken from
    Ref.\,\cite{San14}. The location of the $3/2^-_2$ state is indicated by a
    red cross ($\sigma_{part.} \approx 0.01$\,mb). In addition, the partial
    cross sections relative to the inclusive cross section
    $\sigma_{\mathrm{inc.}}$ are shown, see second axis. Only statistical
    uncertainties are given. No reduction factor $R_s$ has been applied for
    the comparison. See text for further details.}} 
 \end{figure}

Partial cross sections to individual final excited states in $^{55}$Ni and
$^{55}$Co, feeding-corrected where possible, are presented in
Figs.\,\ref{fig:partials}\,{\bf (a)}, {\bf (b)} along with the corresponding
predictions of calculations, {\bf (c)}, {\bf (d)}, combining shell-model
spectroscopic factors with eikonal reaction theory\,\cite{Tos01} following the
approach outlined in\,\cite{Gad08b,Tos14}. As input for the cross-section calculations, the valence-nucleon radial wavefunctions were calculated in a Woods-Saxon-plus-spin-orbit
potential, the geometry of which is constrained by Hartree-Fock calculations using the SkX Skyrme interaction\,\cite{Bro98}.
Shell-model calculations in the full $pf$ shell using the GXPF1A-cd-$pn$
Hamiltonian were used to compute the spectroscopic factors $C^2S(J^{\pi})$ between the $^{56}$Ni ground state and final states with $J^{\pi}$ in $^{55}$Co and $^{55}$Ni, which enter the knockout cross sections. 
GXPF1A is the isospin
conserving part as obtained in\,\cite{Hon02, Hon04, Hon05}. The charge-dependent
(cd) Hamiltonian from\,\cite{Orm89} was added. The isotensor part of this Hamiltonian does not change the wavefunctions for
states with $T=1/2$. The total wavefunctions were calculated in a proton-neutron basis
($pn$). For these shell-model calculations, the computer code NuShellX was utilized\,\cite{NuX}. 
For further details on the calculation of the theoretical cross
sections, see the supplemental material\,\cite{suppl}. In addition to the absolute values, partial cross sections
$\sigma_{\mathrm{part.}}$ relative to the inclusive cross section
$\sigma_{\mathrm{inc.}}$ are shown in Fig.\,\ref{fig:partials}. The ground-state partial cross sections
obtained from subtraction are 29.1(7)\,mb (77(2)\,$\%$) in $^{55}$Ni and
80(5)\,mb (63(4)\,$\%$) in $^{55}$Co. Those values are upper limits
only due to the possibility of missed, weak feeding transitions. 

\begin{figure}[t]
\centering
\includegraphics[width=1\linewidth]{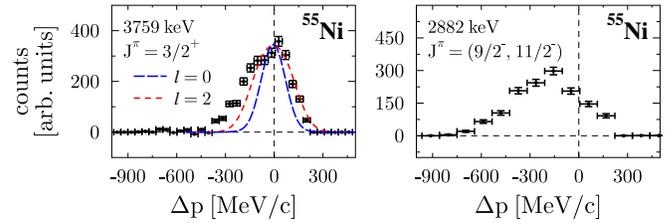}
\caption{\label{fig:parmom}{(color online) Parallel momentum distributions
    measured for the $J^{\pi}= 3/2^+$ and $(9/2^-,11/2^-)$ states in $^{55}$Ni. 
    For the $J^{\pi} = 3/2^+$ state, the predictions of the
    eikonal theory calculated at a mid-target energy of 85.9 MeV/u and folded
    with the momentum distribution of the unreacted beam passing through the
    target are shown with dashed lines. The parallel momentum distribution of the 2882\,keV state was obtained by gating on the high-energy part of the doublet seen in Fig.\,\ref{fig:spectrum}. For both distributions, background contributions were subtracted. Very similar distributions were observed for $^{55}$Co. See text for further discussion.}} 
\end{figure}

For the states
observed in this work, the MED are shown for completeness in the lower panel of
Fig.\,\ref{fig:level} in comparison to the shell-model predictions with the
GXPF1A-cd-$pn$ Hamiltonian ($pf$ states only). The theoretical MED differ by
50-80\,keV. The negative values of the MED relative to the $A=55$ ground states
might be interpreted in terms of an increase of the mean nuclear radii
of the excited states relative to that of the ground state 
due to the increased occupancy of the $2p_{3/2}$ orbital and the connected contribution to the MED through changes in the bulk Coulomb energy from the difference in $Z$ between the mirrors\,\cite{Zuk02, Ben15}. This corresponds to the influence of the monopole radial term on the MED\,\cite{Ben15}.
An MED of -100\,keV corresponds to a 1.0\,$\%$ increase in
the radius. This effect of an increased $2p_{3/2}$ occupancy on the charge radius is similar to the isotope shift of 1.2\,$\%$
observed between $^{56}$Fe and $^{54}$Fe~\cite{Min16}.
The MED for the $1/2^+$ and $3/2^+$ states relative to the $7/2^-$ ground state
also show that the addition of the relativistic spin-orbit correction
of order +100 to +200\,keV (see Table\,IV in\,\cite{Nol69}) as well as the correction for the Coulomb energy stored in a single-proton orbital are required\,\cite{Zuk02}, which corresponds to the influence of the monopole single-particle term on the MED\,\cite{Ben15}. In both nuclei, the $1/2^+$ and
$3/2^+$ states are comparably strongly populated (see Figs.\,\ref{fig:partials}\,{\bf (a)} and {\bf (b)}). These states are expected to have significant contributions to their wavefunction from the $sd$ orbitals below the $N=Z=28$ shell closure and have been previously discussed in\,\cite{San14}, where the $1/2^+$ state in $^{55}$Ni was also strongly populated. The $3/2^+_1$ state had not been unambiguously identified\,\cite{ENSDF, San14}. The measured parallel momentum distributions, see Fig.\,\ref{fig:parmom} for $^{55}$Ni, support a $3/2^+$ assignment
based on the clear observation of a nucleon knockout from an $l = 2$ orbital.  

It is interesting to note that the possible $\left( 9/2^-, 11/2^- \right)$
doublet of the $(2^+_1(\nuc{56}{Ni}) \otimes 1f_{7/2}^{-1})$ multiplet is weakly
populated in this work, 1.42(10)\,mb (3.7(3)\,$\%$) in $^{55}$Ni and
8.2(3)\,mb (6.5(2)\,$\%$) in $^{55}$Co, respectively. The population of these states cannot proceed
by a one-step knockout process from the $^{56}$Ni ground state. The population of such complex configurations
has been reported before, possible due to the knockout from the excited $2^+_1$
state of the projectile (see the discussion in\,\cite{Str14, Mut16}). It has
been speculated in previous studies that such indirect reaction mechanisms result in downshifts observed for some parallel-momentum
distributions\,\cite{Str14, Mut16}. As is shown in Fig.\,\ref{fig:parmom},
the distributions for those states are indeed shifted to lower momenta while the
simpler configurations such as the main fragment of the $1d_{3/2}$ state line up
as expected from the eikonal theory. The structure assignment is supported by the
observation of $B(E2;\left( 9/2^-, 11/2^- \right) \rightarrow 7/2^-_1 )$ values
simlar to the $B(E2;2^+_1 \rightarrow 0^+_1)$ of
$^{56}$Ni\,\cite{Yur04, ENSDF}. The present shell-model calculations predict
large spectroscopic factors between the $2^+_1$ state of $^{56}$Ni and the $(9/2^-, 11/2^-)$ states in the $A = 55$ nuclei of $C^2S = 1.62$ and  $1.12$,
respectively. Theoretically, these core-coupled states are located at energies of about
2.8\,MeV.  

We note that a discussion of the reduction factor
$R_s=\sigma_{\mathrm{exp.}}/\sigma_{\mathrm{th.}}$, reported for a large body of consistently
analyzed knockout reactions\,\cite{Gad08b,Tos14}, is not very instructive here
as knockout from the $sd$ shell is observed, which is outside of the
model space employed by the present shell-model calculations. Nevertheless, we
can make a consistency argument. The theoretically expected
inclusive cross sections, including the $1/2^+$ and $3/2^+$ states with
spectroscopic factors from Ref.\,\cite{San14}, are 85\,mb in $^{55}$Ni and
101\,mb in $^{55}$Co. For \nuc{55}{Ni} this gives a reduction
factor of $R_s = 0.45(4)$ ($\Delta S = 10.5$\,MeV), consistent with expectations
from~\cite{Gad08b,Tos14}. For the slightly more deeply-bound \nuc{55}{Co} ($S_p
>$~5~MeV), we expect that more bound $sd$-shell strength has to be included. Based
on \nuc{57}{Co}\,\cite{Rei87}, the $2s_{1/2}$ strength may be fully exhausted and 75\% of the
$1d_{3/2}$ strength may be found below 5.2\,MeV. Assuming in addition a
spectroscopic factor of 1 for the bound $1d_{5/2}$ strength, and subtracting the
cross section of the indirect contribution identified above, leads to a reduction
factor of 0.93(13) consistent with\,\cite{Gad08b,Tos14}.       

Besides the slightly stronger relative population of the
$5/2^-_1$, $7/2^-_2$, $(9/2^-,11/2^-)$, and $1/2^-_2$ states in $^{55}$Co, the
fragmentation of the spectroscopic strengths between the two lowest-lying $3/2^-$ states is very
different (see Fig.\,\ref{fig:partials}\,{\bf (a)} and {\bf (b)}). One has to consider that this difference may be caused by unobserved feeding (\nuc{55}{Co} is slightly more bound than \nuc{55}{Ni} and will consequently have more bound excited states). For instance,
the unplaced 1352\,keV transition, if feeding the 2566\,keV level, would
decrease its direct partial cross section by $\sim$\,33\,$\%$. The $\gamma$-ray yields 
needed for resolved transitions over an energy range from 0.5\,MeV to 3\,MeV to
obtain comparable cross sections for the $3/2^-$ states due to unaccounted feeding were estimated. If
collected in a single or even two transitions, all of those feeders should have been
identified in the $\gamma$-ray singles spectra. If this asymmetry was indeed
caused by different feeding, the needed strength would have to be fragmented over
multiple transitions which all have to be below the detection limit of the present
measurement. It should be mentioned that the number of levels observed to feed the
$3/2^-_2$ state in $^{55}$Co is larger than in $^{55}$Ni (compare
Fig.\,\ref{fig:level}). Still, after subtraction, its partial cross section is larger.

The observed asymmetry in the partial cross sections is theoretically not expected for the $3/2^-$ states (compare
Fig.\,\ref{fig:partials}\,{\bf (c)} and {\bf (d)}). Therefore, spectroscopic factors $C^2S$ for the one
proton transfer from $^{54}$Fe
(ground state) to $^{55}$Co ($J^{\pi} = 3/2^-_i$) were also calculated and
compared to the data from Table\,1 of \cite{For77}. To obtain agreement between
the shell-model spectroscopic factors and the experimentally determined ratio
of $C^2S(3/2^-_1)/C^2S(3/2^-_2)=1.54(22)$ in $^{55}$Co, derived as the average from several
$^{54}$Fe to $^{55}$Co transfer reactions\,\cite{For77}, mixing amplitudes of
$\alpha = 0.995^{+0.003}_{-0.004}$ ($\alpha^2 = 0.990^{+0.005}_{-0.010}$) and $\beta = -0.10^{+0.04}_{-0.03}$ ($\beta^2 = 0.010^{+0.010}_{-0.005}$)
result in a model with two unperturbed $3/2^-_j$ ($j = I,II)$ states. The wavefunctions of the mixed states are then given by: 

\begin{align*}
|3/2^-_1 \rangle &= \alpha |3/2^-_I \rangle + \beta |3/2^-_{II} \rangle \\
|3/2^-_2 \rangle &= -\beta |3/2^-_I \rangle + \alpha |3/2^-_{II} \rangle
\end{align*}

Without introducing this 0.5\,$\%$ to 2\,$\%$
mixing, the ratio between the corresponding shell-model spectroscopic factors would have been
2.35 ($C^2S(3/2^-_1) = 1.22$, $C^2S(3/2^-_2) = 0.52$). Applying the same mixing to the one-proton knockout from the $^{56}$Ni
ground state to $^{55}$Co leads to spectroscopic factors of $0.181^{+0.002}_{-0.003}$ for $3/2^-_1$ and
$0.006^{+0.003}_{-0.002}$ for $3/2^-_2$ (0.186 and 0.001 without mixing), respectively, not explaining the asymmetry in the partial cross sections  reported here. The emerging contradictory picture prevents conclusions on the role of isospin mixing based on the asymmetry in the partial  cross sections, $\sigma(3/2^-_i)$, and suggests that unobserved feeding in the present data may indeed be a contributor. 

However, independent from the cross-section discussion, the $\gamma$-decay
pattern of the $1/2^+$ state, which is the main feeder of the 
$3/2^-$ levels, is also significantly different in the two mirror nuclei. The phase-space
corrected $R(E1)_{3/2^-_2/3/2^-_1}$ ratios are 2.69(14) and 3.9(3) in $^{55}$Co
and $^{55}$Ni, respectively. To put this into perspective, these numbers mean
that $\sim$\,18\,$\%$ of the feeding-uncorrected $\gamma$-ray yield of the
$3/2^-_2$ state in $^{55}$Co is due to the decay of the $1/2^+$ state while
this contribution is $\sim$\,55\,$\%$ in $^{55}$Ni ($\sim$\,50\,$\%$ and
$\sim$\,48$\%$ for the $3/2^-_1$). The adopted lifetime of $\tau = 71^{+260}_{-4}$\,fs\,\cite{ENSDF}
and the newly determined branching ratio in $^{55}$Co allowed to calculate the
reduced $B\left( E1; 1/2^+ \rightarrow 3/2^-_i \right)$ transition strengths to
be $17.2^{+1.2}_{-13.5}$\,mW.u. to the $3/2^-_1$ and $46^{+4}_{-36}$\,mW.u. to
the $3/2^-_2$, respectively. For low-lying $E1$ transitions, such rates are,
despite the large uncertainty of the lifetime, significantly enhanced. As mentioned in the introduction, a clear change of the $E1$-decay behavior of
an excited state between mirror nuclei, as observed here, has been attributed to isospin-mixing
effects in the $A = 35$\,\cite{Ekm04} and $67$\,\cite{Orl09}
mirror pairs.

With the mixing amplitudes determined from the $^{54}$Fe$-{}^{55}$Co transfer data, two solutions for the unpertubed matrix elements $\langle 3/2^-_{I,II} | T(E1) | 1/2^+_1 \rangle$ can be obtained. The uncertainty of the absolute $B(E1)$ strengths due to the lifetime uncertainty is neglected in the following discussion. It will affect both values in the same way and, thus, not change the ratio between them. In the first case, both matrix elements are positive and large leading to $B(E1;1/2^+_1 \rightarrow 3/2^-_{I,II})$ values of $23(3)$\,mW.u. for the first and $40(3)$\,mW.u. for the second unperturbed $3/2^-$ state, respectively. In the second case, where one of the $E1$ matrix elements is negative, $B(E1)$ values of $12(3)$\,mW.u. and $51(3)$\,mW.u. are obtained. In both cases the first $E1$ matrix element is also comparably large. We note that an $E1$ transition between pure $(2s_{1/2})^{-1}(1f_{7/2})^8$ hole and $(1f_{7/2})^{6}(2p_{3/2})^1$ particle configurations for the $1/2^+$ and $3/2^-$, respectively, would be forbidden. Consequently, more complex configurations have to be present to explain the enhanced $E1$ rates. In fact, the relative partial cross
sections of 6.1(3)\,$\%$ in $^{55}$Ni and 6.6(3)\,$\%$ in $^{55}$Co are almost
identical for the $1/2^+$ state (compare Fig.\,\ref{fig:partials}), which suggests a similar structure of the $1/2^+$ state in the mirror pair and supports the hypothesis that the observed change in the $E1$ decay pattern probes the degree of mixing between the two $3/2^-$ states. The amount of isospin mixing needed to explain the $E1$ asymmetry for a low-lying $7/2^-$ level in the $A = 35$ and a $9/2^+$ state in the $A =
67$ mirror nuclei was estimated to be on the order of 1\,$\%$ to 5\,$\%$\,\cite{Ekm04, Pat08, Orl09, Biz12}. In contrast to Refs.\,\cite{Ekm04, Pat08, Orl09, Biz12}, no mixing for the initial and final states but only between the two final states was assumed in the mixing scenario discussed here. Interestingly, the unperturbed $R(E1)_{3/2^-_{II}/3/2^-_I}$ ratio is $4.3^{+1.7}_{-1.1}$ in the second case, i.e. closer to the experimentally observed ratio for $^{55}$Ni. The smaller experimentally observed $R(E1)$ ratio in $^{55}$Co might, thus, point at stronger mixing between the unperturbed $3/2^-$ states than in its mirror $^{55}$Ni.

In conclusion, we have performed the first mirrored one-nucleon knockout
reactions on the self-conjugate nucleus $^{56}$Ni leading to the mirror pair
(\nuc{55}{Ni},\nuc{55}{Co}). From in-beam $\gamma$-ray spectroscopy, partial
cross sections were determined and the $\gamma$-decay properties were studied for a
number of excited states in $^{55}$Ni and $^{55}$Co. Several states carrying
single-particle strength were populated in the $A = 55$ mirror pair. The
fragments of the $2s_{1/2}$ and $1d_{3/2}$ hole-states carry significant cross
sections in both nuclei, emphasizing the necessity to include $sd$ orbitals
for the description of nuclei in this region. Small cross sections to a potential
doublet of core-coupled states $(2^+_1({}^{56}\mathrm{Ni}) \otimes 1f_{7/2}^{-1})$,  $(9/2^-,11/2^-)$, were also observed,
together with a telltale downshift in their parallel
momentum distributions, indicative of an indirect reaction pathway. 
A pronounced cross section asymmetry for the two lowest-lying $3/2^-$ states as well as a clear change in the $E1$ decay pattern
of the $1/2^+_1$ level feeding the $3/2^-$ states were discussed. The high degree of mixing, that would be needed to explain the cross-section asymmetry in a
two-level approach, cannot be reconciled with a comparison of data on the transfer from $^{54}$Fe to $^{55}$Co to the corresponding shell-model calculations preventing conclusions on the role of isospin mixing. The change in the $E1$ decay pattern, however, hints at stronger mixing between the $3/2^-$ states in $^{55}$Co than in its mirror $^{55}$Ni and reveals an unexpected mirror asymmetry close to the nominally doubly-magic $N=Z$ nucleus $^{56}$Ni.

\begin{acknowledgments}
This work was supported by the National Science Foundation under Grant No. PHY-1565546 (NSCL) and PHY-1811855, the DOE National Nuclear Security Administration through the Nuclear Science and Security Consortium under Award No. DE-NA0003180 as well as by the UK Science and Technology Facilities Council (STFC) through Research Grant No. ST/F005314/1, ST/P003885/1 and GR/T18486/01. M.S. wants to thank S.\,Giuliani, G.\,Potel and J.\,Rotureau for helpful discussions.
\end{acknowledgments}

\bibliography{56Ni}

\end{document}